  \documentclass[twocolumn,showpacs,preprintnumbers,amsmath,amssymb]{revtex4-2}

\usepackage[colorlinks=true,citecolor=blue,urlcolor=blue,linkcolor=blue,bookmarksopen]{hyperref}
\usepackage{amsfonts}
\usepackage{amsmath}
\usepackage{amssymb}
\usepackage{bbold}
\usepackage{mathrsfs}
\usepackage{graphicx}
\usepackage{dcolumn}
\usepackage{bm}
\usepackage{times}
\usepackage{comment}
\usepackage{latexsym}
\usepackage{epsf}
\usepackage{float}
\usepackage{natbib}
\usepackage{bm}
\usepackage[export]{adjustbox}
\usepackage[makeroom]{cancel}
\usepackage[english]{babel}
\usepackage[T1]{fontenc}
\usepackage{wasysym}
\usepackage[tight, FIGTOPCAP, hang, raggedright, nooneline]{subfigure}
\usepackage[dvipsnames]{xcolor}
\usepackage{epsfig}
\usepackage{lipsum}
\usepackage{enumitem}
\usepackage{mathtools}
\usepackage{footnote}
\usepackage{ragged2e}
\usepackage{cleveref}

\usepackage{perpage}
\MakePerPage{footnote}

\DeclarePairedDelimiter\ket{\lvert}{\rangle}
\DeclarePairedDelimiterX\braket[2]{\langle}{\rangle}{#1 \delimsize\vert #2}
\renewcommand{\vec}[1]{\boldsymbol{#1}}

\begin{document}

\title{How many supercells are required to achieve unconventional light confinement effects in moiré photonic lattices?}

\author{Chirine Saadi$^1$}
    \email{chirine.saadi@ec-lyon.fr}
\author{Hai Son Nguyen$^{1}$}
\author{S\'ebastien Cueff$^1$}
\author{Lydie Ferrier$^1$}
\author{Xavier Letartre$^1$}
\author{S\'egol\`ene Callard$^1$}
    \email{segolene.callard@ec-lyon.fr}
    
\affiliation{$^1$Univ Lyon, Ecole Centrale de Lyon, CNRS, INSA Lyon, Universit\'e Claude Bernard Lyon 1, CPE Lyon, CNRS, INL, UMR5270, 69130 Ecully, France} 

\date{\today}	
\pacs{}

\begin{abstract}
Moiré structures are receiving increasing attention in nanophotonics as they support intriguing optical phenomena. In the so-called “magic configuration”, one-dimensional moirés give rise to fully dispersionless energy bands known as “flatbands", where the light is tightly localized within \textbf{each} supercell of the periodic moiré. The goal of this investigation is to determine to what extent the confinement of light, observed in periodic structures, is preserved in microcavities of finite size. Here we analyze the optical response of finite moiré structures consisting of one, two, or more supercells of 1D moiré. Our calculations reveal that for single-supercell cavity, the magic configuration does not impact the electric field confinement at the wavelength of the flat band modes. 
However, when three or more supercells are connected, we show that the coupling between supercells is canceled at the “magic configuration,” resulting in highly confined modes with a quality factor greater than $10^6$ 
and exhibiting the characteristics of a quasi-bound state in the continuum where optical losses are eliminated through a destructive interference process.
\end{abstract}

\maketitle

\section{Introduction}
The discovery of moiré flatbands in twisted bilayer graphene (TBG) at "magic angles"\cite{Bistritzer2011} sparked the birth of "twistronics" in two-dimensional (2D) materials\cite{Hennighausen2021}. This has revealed unconventional superconductivity\cite{Cao2018}, highly correlated electronic states with non-trivial topology\cite{bultinck2020mechanism,Serlin2020,Bhowmik2022}, and moiré excitons\cite{jin2019observation,Alexeev2019} . Inspired by these breakthroughs in solid state physics, the photonic field has recently seen a surge of research in optical moiré superlattice implemented in monolayer\cite{wang2020localization,fu2020optical,Mao2021, Talukdar2022,Yang2022} and bilayer\cite{Lou2021,Dong2021,Tang2021,nguyen2022magic,Hong2022,Guan2023} photonic crystals. Analogous to magic angle TBG, photonic flatbands with zero light group velocities and sharp photonic local densities of states (LDOS) can emerge at photonic magic angles of twisted 2D photonic crystals\cite{Dong2021,Tang2021}.  Moreover, numerous effects have been demonstrated on twisted moiré superlattices, such as light localization\cite{wang2020localization, Yang2022}, chiral polarization\cite{Dong2021}, lasing action\cite{Mao2021} and far-field couplings\cite{Guan2023}.

Extending beyond the "twist" configuration, perfect flatbands can be generated in one-dimensional (1D) moiré bilayer photonic crystals consisting of two mismatched 1D gratings separated by "magic distances"\cite{nguyen2022magic}.  Intriguingly, the formation of 1D moiré minibands, thoroughly explained by a microscopic model encompassing inter- and intra-layer coupling mechanisms, can be captured by a straightforward tight-binding model of a mono-atomic chain of moiré supercells. At magic distances, photon tunneling between trapped moiré Wannier states is canceled due to the accidental destructive interference between inter- and intra-layer couplings, resulting in flatband formation and unconventional localization of photonic states within a single moiré period \cite{nguyen2022magic}. This configuration holds great potential for enhancing light-matter interactions. For example, leveraging the high LDOS of these magic-distance flatbands when coupled to active materials, a remarkable enhancement factor of $5.10^5$ for second-harmonic generation has been proposed \cite{Hong2022}. Another attractive application of this magic configuration for light-matter interaction is the engineering of photonic cavities with a high figure of merit $Q/V$ ratio where $Q$ represents the quality factor, and $V$ denotes the modal volume. In this regard, a recent study by Tang et al.\cite{tang2022chip} suggested that magic flatbands in 1D moiré bilayer photonic crystals outperform traditional photonic crystal structures such as defect photonic crystal cavities and BIC photonic crystals, thanks to their extremely high $Q$, and presumably small $V$ derived from the unconventional localization regime. 
However, this assumption is not straightforward, as most simulations for Moiré bilayers are employing periodic conditions, rendering the modeled structures, as well as their modal volume $V$, infinite. So far, it is unclear whether unconventional light confinement effects would be preserved in finite-sized Moiré superlattices. In that context, exploring light confinement within finite moiré cavities and understanding its relationship with the magical configurations observed in infinite bilayer photonic crystal moirés remain open questions.

In this study, we delve into the unexplored domain of light-confinement regimes in finite moiré cavities by systematically analyzing various models derived from 1D moiré bilayer photonic crystals superlattices. These models comprise either one isolated supercell or two, to several connected moiré supercells.
Hence, our investigation centers around the quality factor and modal volume of finite moiré cavities, as well as the extent to which light localization observed at magic distance in infinite structures is maintained within cavities of finite dimensions. 
Remarkably, we observe that a non-Hermitian Hamiltonian, which incorporates photon tunneling between moiré supercells and accounts for radiative losses at the edges of the cavity, accurately describes the physics of the system and aligns flawlessly with numerical simulations using the finite element method. This intriguing finding highlights the effectiveness of the non-Hermitian model in capturing the behavior of the system. Our research demonstrates that despite achieving a reasonable level of light confinement within a 4 µm isolated single-supercell cavity, with a mode exhibiting a quality factor of approximately $10^3$, the magic configuration does not influence this confinement in any significant way. 
In contrast, when considering a triple-supercell moiré cavity, the confinement is notably strengthened when utilizing magic distance configurations. This leads to the creation of a quasi-bound state in the continuum (quasi-BIC) with a remarkably high quality factor of approximately $Q\sim 10^7$. This quasi-BIC is tightly confined within the central 4 µm supercell. The emergence of the moiré quasi-BIC can be explained by the isolation of the middle supercell, where tunneling cancellation at magic distance prevents radiative losses at the lateral boundaries of the cavity. Compared to more trivial forms of confinement using potential barriers, these findings open up exciting opportunities for leveraging the magic configurations of moiré bilayer photonic crystals to explore novel regimes of cavity electrodynamics, nonlinear physics, sensing, and lasing. 

\section{Moiré Bilayer Photonic crystal}
\begin{figure}[ht!]
\centering
\includegraphics[ width=\linewidth]{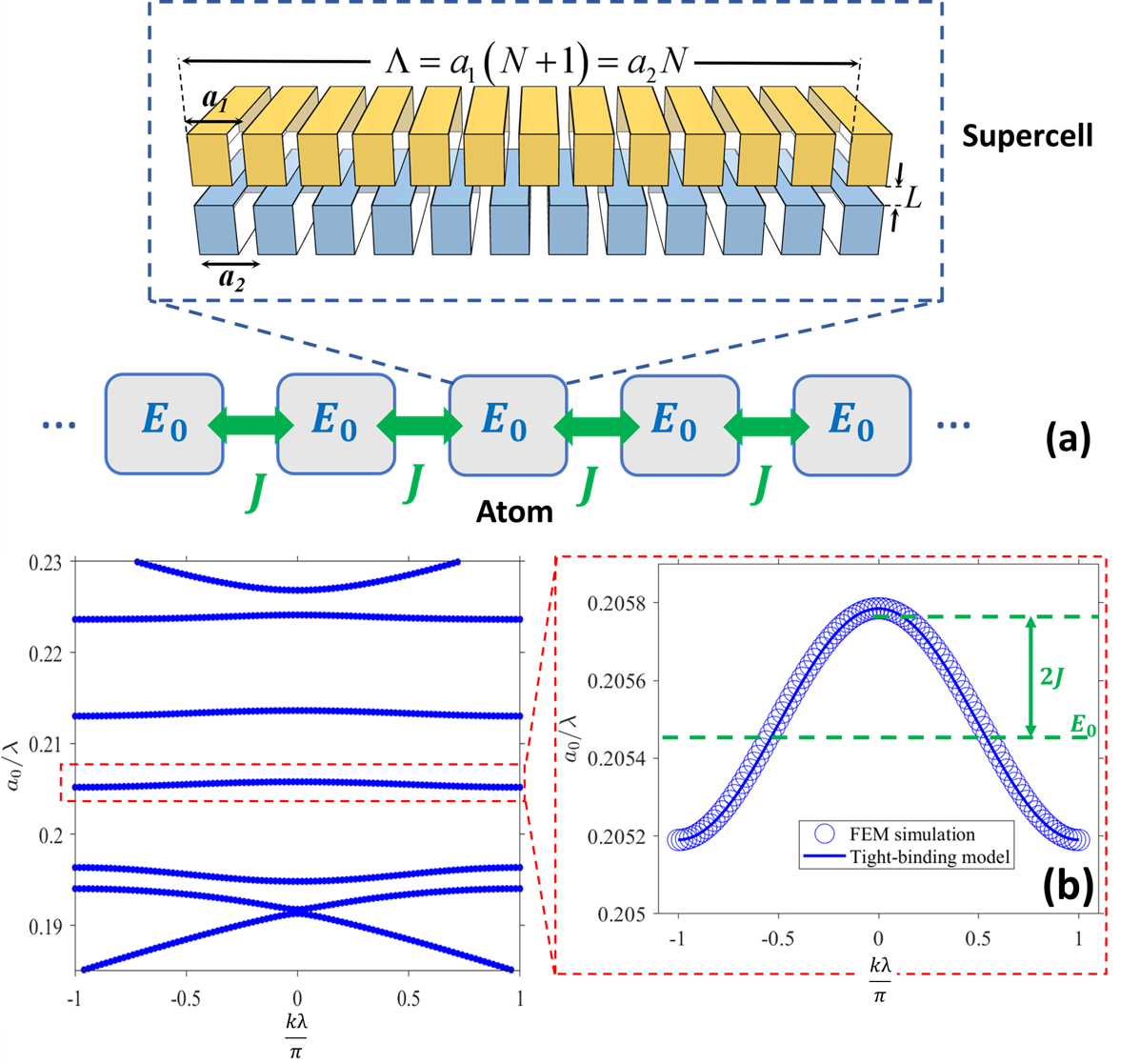}
\caption{\label{figure1}(a) Sketch of a chain of moiré meta-atoms having an energy $E_0$ and a nearest-neighbour coupling strength $J$. Each meta-atom represents one supercell of the 1D moiré structure. (b) Illustration of a band diagram of 1D moiré at non-magic distance ($L=0.1$) computed by FEM. simulations. Low energy band: fit of the FEM simulations of 1D moiré (circle) by the  tight binding model (solid line) of \eqref{eq:refname}. For this example of non-magic distance, we extract  $E_0\sim0.2055$ and $J\sim1.49\, 10^{-4}$. Here all the energies are normalized by $hc/a_0$.}
\label{fig:false-color}
\end{figure}

In this section, we present a comprehensive revisit and novel insights into the results obtained on 1D periodic moiré structures composed of two superimposed infinite gratings separate by a distance $L$ \cite{nguyen2022magic}. More precisely, we investigate the photonic band structure and the quality factor of the modes of the infinite moiré as a function of $L$. The 1D moiré consists of two mismatched silicon (n=3.54) photonic crystal slabs of thickness $h$ and filling factor $\kappa$, surrounded by air and presenting slightly different periods $a_1$ and $a_2$, with an average period of $a_0$=$\frac{a_1 + a_2}{2}$. The periods are chosen to satisfy the commensurate condition $a_1/a_2=N/(N+1)$, allowing to generate a periodic 1D moiré of period $\Lambda=(N+1)a_1=Na_2$. The parameters of the gratings ($a_0=300$, $h=180nm$ and $\kappa=0.8$) are chosen to obtain flat bands in the telecom wavelength range around $\lambda\approx 1.5 \mu m$ . For the sake of illustration and without loss of generality, $N$ is fixed to $13$ ( $\Lambda\approx 4 \mu m$). The band structures are computed with a finite element method (FEM) using the COMSOL Multiphysics software. In coherence with previous results \cite{nguyen2022magic}, two ultra-flat bands appear for peculiar values of $L$ called "magic distances". In this study, we focus on the lower energy band (fig.\ref{figure1}.b) also referred as hole-like band. We show that for this band (fig.\ref{figure1}.b), and for each $L$, the computed energy profile between $k=0$ and $k=\frac{\pi}{\Lambda}$ (respectively the $\Gamma$ and the $X$ points of the first Brillouin zone) can be described by the following equation, where the normalized energy is given by :
\begin{equation}
E(k) =\frac{a_0}{\lambda(k) }= E_0+2J\cos(\Lambda k)
\label{eq:1}
\end{equation}
This behavior is explained in the framework of a tight-binding model \cite{nguyen2022magic} in which the moiré is modeled by a periodic chain of meta-atoms where a meta-atom represents a moiré supercell, the distance between two meta-atoms is $\Lambda$ and each meta-atom of energy $E_0$ is coupled to its next neighbors by a tunneling rate $J$ (fig.\ref{figure1}.a). Both $E_0$ and $J$ depend on $L$. For each $L$, the simulations allow to determine the average energy of the band  within the whole momentum space along with its bandwidth (see supplementary information). When $L$ reaches “magic distances”, for $L/a_0=0.027$ and $L/a_0=0.187$, calculations show that the bandwidth of the band achieves its lowest value: accordingly, at these distances, the coupling between each supercell cancels ($J=0$) as if each supercell was isolated from its neighbors. Secondly, our calculations also show that the quality factors of the flat-band mode remain above $10^9$ within the entire momentum space (table S.1 in supplemental information).  This indicates that the structures exhibit extremely low losses. It is important to note that these calculations were performed considering the periodic structure, allowing for the determination of the values of  $E_0(L)$ and $J(L)$. These values are then used as input parameters for the model.

\section{Moiré microcavities}
\subsection{Single-supercell cavity and double-supercell cavity}
\begin{figure*}[ht!]
\centering
\includegraphics[width=\linewidth]{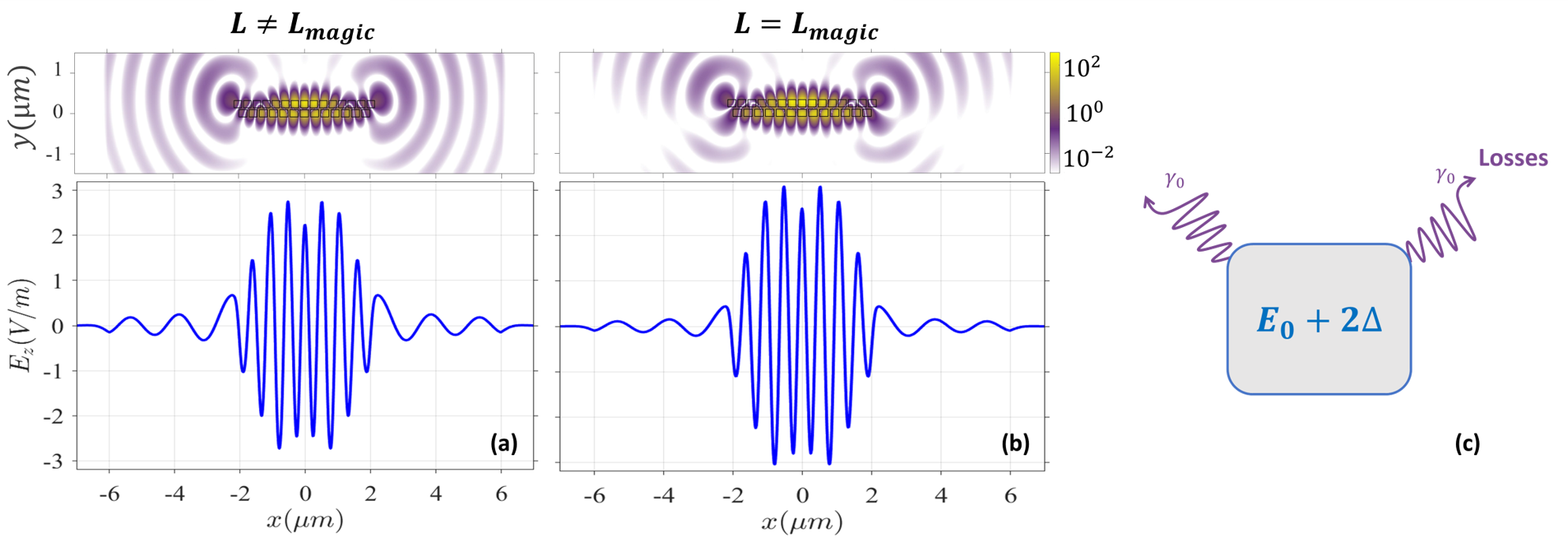}
\caption{\label{figure2}(a,b) Localization of the electric field mode and its corresponding electric field distribution localized in a single-supercell cavity (a) non - magic distance for $L/a_0 =0.152$ and (b) at magic distance $L/a_0 =0.187$. (c) The corresponding model composed by one meta-atom that have an energy $E_0+2\Delta$ shifted with respect to $E_0$.The losses $\gamma_0$ are occurring at the extremities of the structure.}
\label{fig:false-color}
\end{figure*} 

\begin{figure}[ht!]
\centering
{\includegraphics[width=0.9\linewidth]{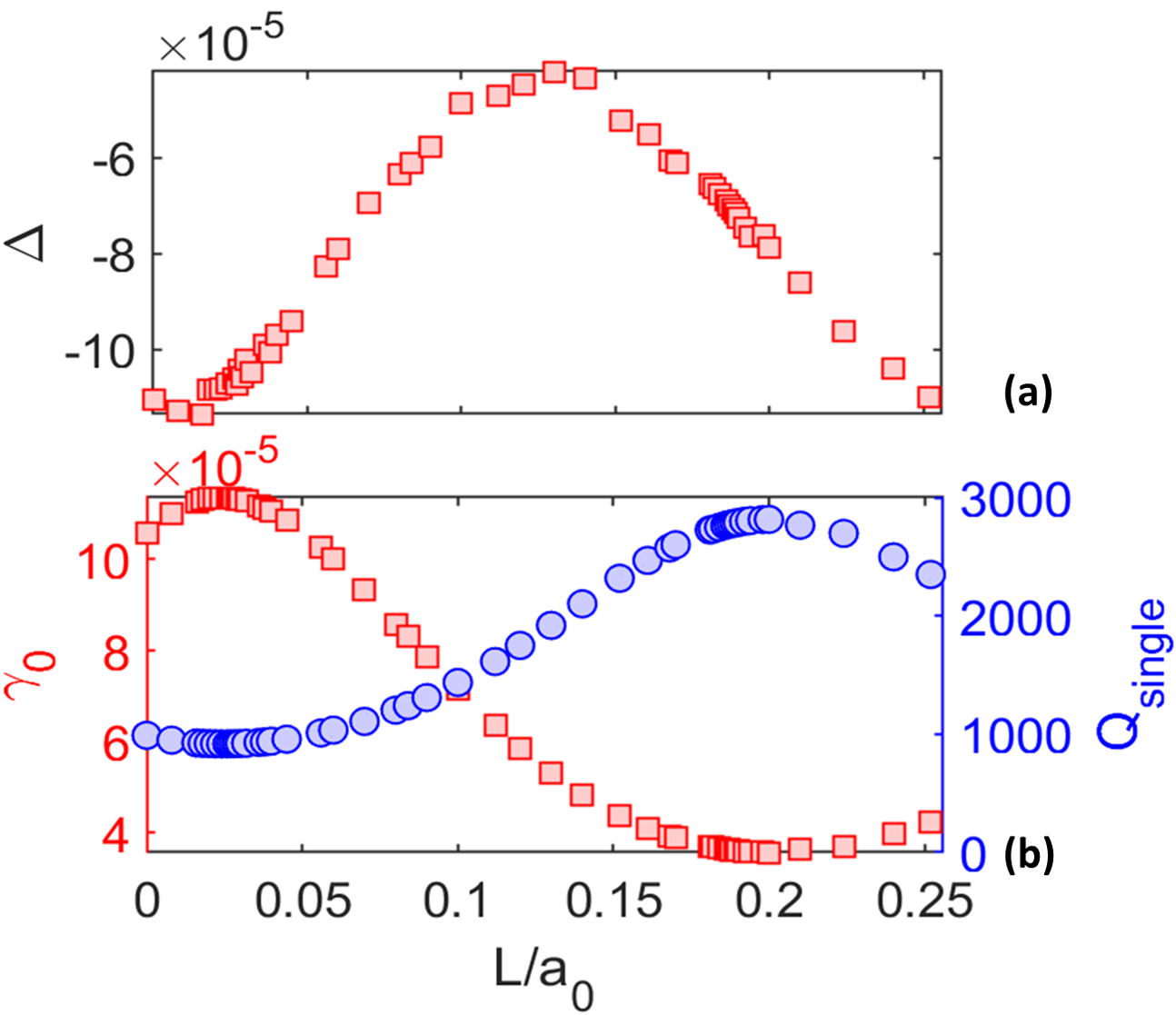}}
\caption{\label{figure3} (a) Computed shifted energy, (b)losses (squares) and quality factors (circles) of a single-supercell cavity according to different values of $L$.}
\label{fig:false-color1}
\end{figure}
To investigate the properties of a true isolated moiré supercell, we considered a structure formed by a unique moiré pattern (fig.\ref{figure2}.c). It corresponds to a single meta-atom of the periodic structure modeled in the previous section. The modal properties of the so-formed cavity are calculated as a function of $L$ using COMSOL. Firstly, for each $L$, a stationary mode is found at $E^{single}$ in the vicinity of the former flat band energy $E_0$. The numerical simulations demonstrate that regardless of the distance $L$, this mode is confined within the supercell, exhibiting a distribution similar to the pattern observed in one supercell of the periodic moiré (fig.\ref{figure2}.a-b). In our model, its energy can be expressed as $E^{single}=E_0+2\Delta$, where $\Delta<0$ is an energy shift that accounts for the change of the boundaries condition at the edges of the supercell when surrounded by free space. $E^{single}$ and $\Delta$ indeed depends on $L$. Secondly, the calculation of the quality factor of this mode shows a dramatic decrease with respect to those of periodic structure ($10^3$ instead of $10^9$). This is because the photons can now escape through the supercell boundaries with free space (fig.\ref{figure2}.c). Assuming that this phenomenon is the principal channel of losses which is confirmed in (fig.\ref{figure2}.a-b) , the global loss rate $\gamma^{single}$ is related to the loss rate at one edge $\gamma_0$ by $\gamma^{single}=2\gamma_0$ (or $2Q^{single}=Q_0$ ).
Once again, numerical simulations allow for the determination of $\Delta$, $\gamma_0$ and $Q_{single}$ as a function of $L$ (fig.\ref{figure3}.a-b). These computations are used as input parameters to finalize the model for our investigation. Note that with one supercell, the magic configuration has no striking effect on the losses: the quality factor of the mode remains approximately around $10^3$ regardless of $L$.

In the case of the double-supercell cavity, also refereed as "diatomic molecule" \cite{nguyen2022magic}, the model is composed of two meta-atoms coupled to each other through the tunneling rate $J$, which has been defined for the infinite structure. Additionally, each meta-atom presents a loss rate of $\gamma_0$ at its interface with free space, defined previously in the single meta-atom case. The model predicts that the hole-like bound state in the microcavity separates into bonding and antibonding states. Indeed, the numerical simulations show that these states are degenerate and equal to $E_0+\Delta$ at the magic distances, when $J$ vanishes. However, they also show that the magic distance has no effect on the quality factors of the bonding and antibonding modes: as for the single-supercell cavity, they remain of the order of $10^3$ regardless of $L$ (see supplemental information).To investigate the impact of the magic distance on light confinement, we plan to connect additional supercells of moiré.

\subsection{Triple-supercell cavity}
\begin{figure}[ht!]
\centering
{\includegraphics[width=\linewidth]{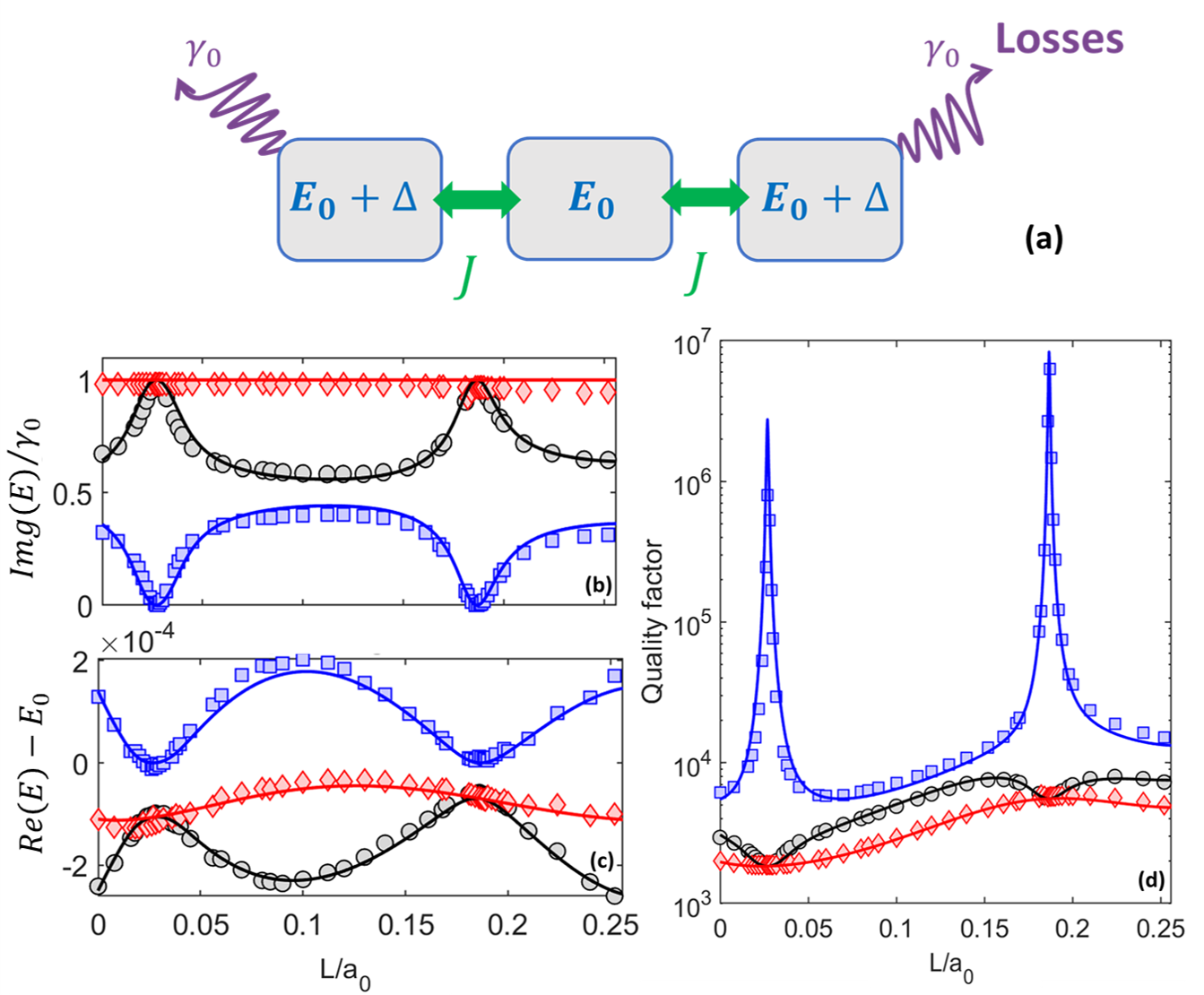}}
\caption{\label{figure4}(a): Sketch of the 1D moiré triple-supercell in terms of atom-like structure. 
(b) Losses(imaginary part of complex energy), (c) Energies (real part of complex energy), and (d) Quality factors of mode 1 (black circles), mode 2 (red diamonds) and mode 3 (blue squares) in function of $L/a_0$ in the triple-supercell cavity. Symbols refer to COMSOL computation and solid lines to the analytical model. The losses and the quality factor variations of the modes demonstrate the emergence of a quasi-BIC at magic distances for $L/a_0 = 0.027$ and $L/a_0 = 0.187$. }
\end{figure}

Based on the previous results, we design a structure made up of three coupled supercells (fig.\ref{figure4}.a). Here, we anticipate that at magic distances the tunneling rate $J$ will vanish, implying that the central meta-atom will be isolated from the two others. As a result, we expect to find a quasi-lossless mode confined in the middle supercell. With the results of the former simulations, we build an semi-analytical model to describe the triple-supercell system. As the middle meta-atom is surrounded by two other meta-atoms, we assume that, its energy $E_M$ will be equal to the energy of the meta-atom $E_0$ in the 1D periodic moiré. Conversely, the left and right meta-atoms are surrounded by one meta-atom and free space. Hence, as for the double-supercell cavity, their energy will undergo a shift $E_{R,L}=E_0 +\Delta$. As before, the shift in energy $\Delta$ accounts for the effect of the interface with free space. In our model, the losses are concentrated at the edges of the structure and are identical to the losses $\gamma_0$ defined and computed for the single-supercell cavity. The three supercells are still coupled with the tunneling rate $J$. All the parameters $E_0$, $J$, $\Delta$ and $\gamma_0$ are functions of $L$ and have been determined by the previous computations (1D periodic moiré and single-supercell).

Using the basis of the localized states in each meta-atom, namely $\ket{L}$ for the mode localized in the left meta-atom, $\ket{M}$ for the middle meta-atom and $\ket{R}$ for the right meta-atom, the Hamiltonian of the system can be described by the following expression:
\begin{equation} H=
\begin{pmatrix}
   E_0+\Delta  & J & 0\\
   J & E_0 & J \\
   0 & J & E_0+\Delta
\end{pmatrix}
+i \begin{pmatrix}
   \gamma_0  & 0 & 0 \\
   0 & 0 & 0 \\
   0 & 0 & \gamma_0
\end{pmatrix}
\label{eq:H}
\end{equation}
From the diagonalization of $H$, we obtain the eigenvalues, $E_1$, $E_2$ and $E_3$ of the system:
\begin{align}
 \begin{split}
E_1 &= E_0+\frac{\Delta}{2}+i\frac{\gamma_0}{2}+\sqrt{\left(\frac{\Delta}{2}+i\frac{\gamma_0}{2}\right)^2 + 2J^2},\\
E_2 &= E_0+\Delta+i\gamma_0,\\
E_3 &= E_0+\frac{\Delta}{2}+i\frac{\gamma_0}{2}-\sqrt{\left(\frac{\Delta}{2}+i\frac{\gamma_0}{2}\right)^2 + 2J^2}\label{eq:eigenvalues}.
\end{split}
\end{align}
The corresponding eigenvectors, also called mode 1, mode 2 and mode 3, are given by:
\begin{align}
 \begin{split}
\ket{\Psi_1} &\propto\left[-1,\frac{\Delta+i\gamma_0}{2J}-\sqrt{\left(\frac{\Delta+i\gamma_0}{2J}\right)^2+ 2},-1\right]
,\\
\ket{\Psi_2} &\propto\left[-1,0,1\right]
,\\
\ket{\Psi_3} &\propto\left[-1,\frac{\Delta+i\gamma_0}{2J}+\sqrt{\left(\frac{\Delta+i\gamma_0}{2J}\right)^2+ 2},-1\right]
\label{eq:eigenstates}
\end{split}
\end{align}
At magic configurations ($J=0$), the eigenvalues of \eqref{eq:eigenvalues} are simplified into: 
\begin{align}
\begin{split}
E_1^{magic} &= E_0+\Delta+i\gamma_0,\\
E_2^{magic} &= E_0+\Delta+i\gamma_0,\\
E_3^{magic} &= E_0,
\label{eq:eigenvalues_magics}
\end{split}
\end{align}
The corresponding eigenvectors \eqref{eq:eigenstates} at magic configurations become:
\begin{align}
\begin{split}
\ket{\Psi_1}^{magic}&=\frac{1}{\sqrt{2}}\left[-1,0,-1\right],\\
\ket{\Psi_2}^{magic}&=\frac{1}{\sqrt{2}}\left[-1,0,1\right],\\
\ket{\Psi_3}^{magic}&=\left[0,1,0\right].
\label{eq:eigenstates_magics}
\end{split}
\end{align}

 As expected, the semi-analytical model predicts that at “magic distances”, the mode 3 of eigenvector $\ket{\Psi_3}^{magic}$ is lossless and strictly localized inside the middle supercell. This result is explained by the isolation of the middle cell $\ket{M}$ from the two lossy cells $\ket{L}$ and $\ket{R}$ when the coupling $J$ becomes zero at magic configurations. On the other hand, the two other modes of eigenvectors $\ket{\Psi_1}^{magic}$ and $\ket{\Psi_2}^{magic}$, are symmetric and anti-symmetric combination of $\ket{R}$ and $\ket{L}$ modes. Therefore they are purely localized in the external supercells, thus degenerate at energy $E_0+\Delta$ and  both lossy with the same losses $\gamma_0$. 
 
\begin{figure}[ht!]
\centering{\includegraphics[width=\linewidth]
{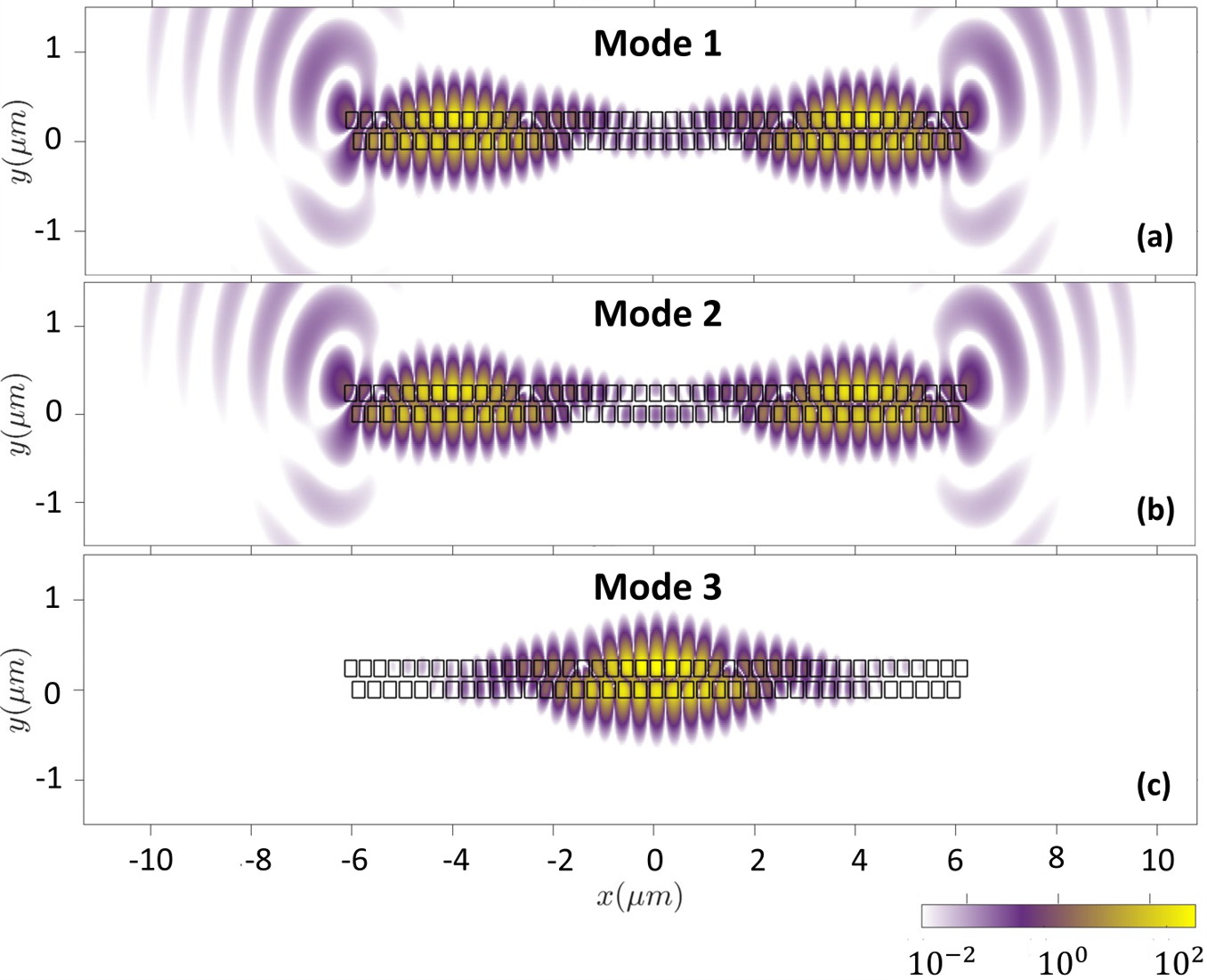}}
\caption{\label{figure5} Amplitude maps ($E_z$) of the modes of a triple-supercell cavity at magic distance $L/a_0=0.187$. (a) Lossy   mode 1, (b) Lossy mode 2, (c) Quasi-BIC mode 3. Modes 1 and 2 are localized on both the right and the left supercells. Mode 3 is confined in the middle supercell and lossless.}
\label{fig:false-color}
\end{figure}
 Numerical FEM computations allow to confirm the above predictions. We first focus on the eigenvalues of the three resulted modes. Figures (\ref{figure4}.b,c) compare the values of the energies/losses of the three modes obtained by FEM simulations with the real/imaginary part of the \eqref{eq:eigenvalues} when scanning the interlayer distance $L$. These results show that the analytical model and the numerical simulations match perfectly with no fitting parameters (we remind that the input parameters for the analytical models are retrieved from the infinite structure and the single-supercell configuration as discussed in the previous sections). Importantly, the numerical simulations also confirm that an ultra-high quality factor, in the order of $10^7$,  is obtained at magic distances, while it remains below $10^4$ for non-magic distances (fig.\ref{figure4}.d).  We note that the finite value of the quality factor at magic configurations is due to small out-of-plane losses that are not considered in the analytical model \eqref{eq:H}. Therefore $\ket{\Psi_3}^{magic}$ is not a perfect BIC but a quasi-BIC state with perfect lateral confinement and almost-perfect vertical confinement.One may implement this losses by adding a diagonal term $i\gamma_\infty$ to \eqref{eq:H} with $\gamma_\infty \ll \gamma_0$ being the negligible  out-of-plane losses. For the analytical fitting in fig.\ref{figure4}.d, we use $\gamma_\infty = \gamma_0/1500$. \\

To check the validity of the analytical model with respect to the eigenvectors, we investigate the light distribution of the modes in the triple-supercell microcavity at the magic distance $L/a_0=0.187$. The computed intensity maps of the three modes are presented in fig.\ref{figure5}. It shows that  two degenerate modes, mode 1 and mode 2, found respectively at the normalized energies of $E_1=0.20185$ and $E_2=0.20186$  are indeed localized in the external supercells (fig.\ref{figure5}.a-b) which is expected for the eigenvectors $\ket{\Psi_1}^{magic}$ and $\ket{\Psi_2}^{magic}$ according to \eqref{eq:eigenstates_magics}. These are lossy modes presenting both the same quality factor of 5730. Conversely, the third mode, mode 3, is found at $E_3=0.20193$ which is, as expected, the value of $E_0$ at the "magic distance" $L=0.187 a_0$. This mode presents the appearance of a quasi-BIC: it has a high quality factor $Q\sim6.3\times10^6$ and the field distribution is well confined in the middle cell (fig\ref{figure5}.c) which clearly corresponds to the case of mode 3:  $\ket{\Psi_3}^{magic}=\ket{M}$. Note that the light distribution of mode 3 (fig\ref{figure5}.c) is very similar to the light distribution of the mode in the single-supercell cavity (fig\ref{figure2}.a-b) except for the evanescent penetration into the adjacent supercells. The striking difference between these results is the absence of radiating losses of the mode 3. This can be explained by the interference effect of the moiré which, at "magic distance", suppresses the coupling between the supercells, and thus prevents any "contact" of the middle supercell with the lossy interfaces of the two external supercells. We stress that although the coupling is forbidden ($J=0$), the presence of the two external supercells is necessary to cancel the losses: indeed, the study of the single supercell shows that it is not possible to avoid losses with a simple interface with free space (see fig.\ref{figure2}).

\begin{figure}[ht!]
\centering
{\includegraphics[width=\linewidth]{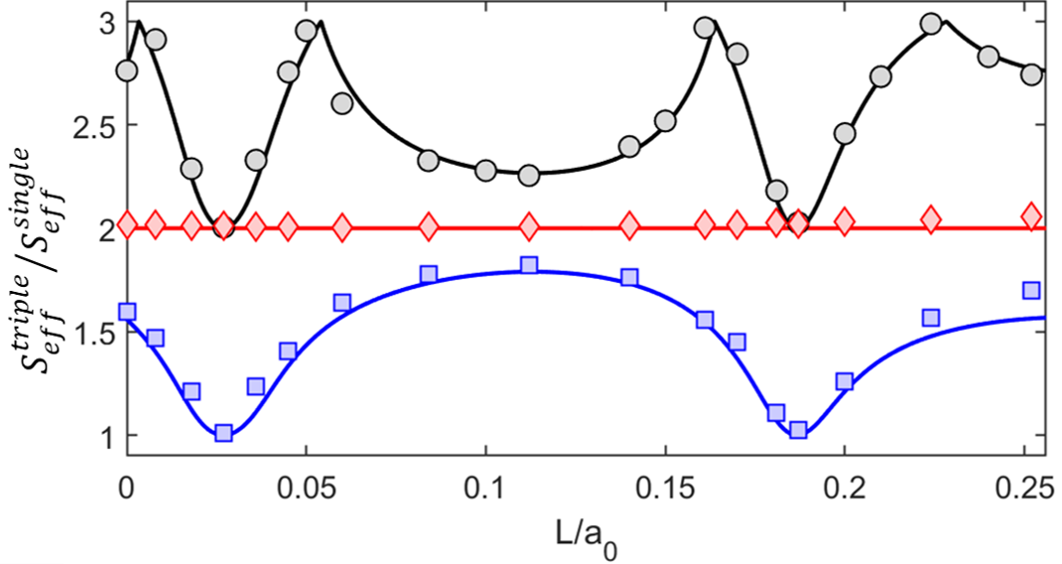}}
\caption{\label{fig6} Ratio of the effective modal surface of mode 1 (black circle), mode 2 (red diamonds) and mode 3 (blue squares) of a triple-supercell cavity with respect of the effective modal surface of the flatband mode in a single-supercell cavity as a function of $L/a_0$. The solid lines correspond to the analytical model and the markers refer to  FDTD simulations.}
\label{fig:false-color3}
\end{figure}

\begin{figure}[ht!]
\centering
{\includegraphics[width=\linewidth]{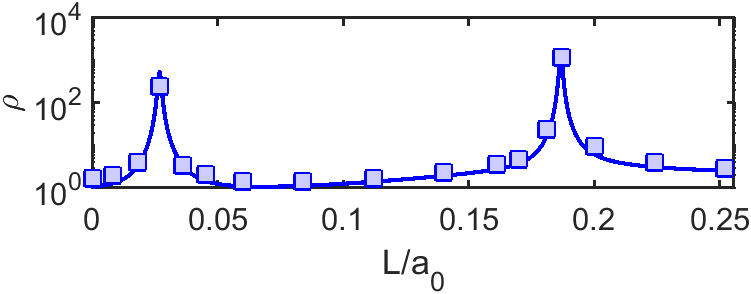}}
\caption{\label{fig7}Variation of the Purcell enhancement ratio  $\rho$ as a function of $L/a_0$. The solid lines correspond to the analytical model and the markers refer to the FDTD simulations.}
\label{fig:false-color}
\end{figure}

To evaluate quantitatively the light confinement in finite moiré microcavities, we employ Finite Difference Time Domain (FDTD) simulations from Rsoft software to compute the effective model surface of eigenmodes in single- and triple-supercell configurations. The effective modal surface characterizes light confinement in 2D systems, playing a role similar to that of the effective modal volume in 3D systems. For an eigenmode of field distribution $\vec{E(x,y)}$, the effective modal surface is defined by:
\begin{equation}
    S_{eff}=\frac{ \iint_S \lvert \vec{E(x,y)} n(x,y) \rvert^2 dxdy}{\max{ \lvert \vec{E(x,y)} n(x,y) \rvert^2 }}.
\end{equation}
Here $\vec{E(x,y)}$ has only $E_z$ component for TE modes, and $n(x,y)$ corresponds to the refractive index distribution of the microcavity. For the single-supercell cavity, the numerical results (see Supplemental Information) show that  the magic distance has no impact on the effective modal surface. 
However, for the triple-supercell cavity, the dependence of the effective surface on the distance $L$ is predicted to be much richer. Indeed, for an eigenvector $\ket{\Psi}=\left[c_L,c_M,c_R\right]$ of the triple-supercell cavity, where $c_L$, $c_M$ and $c_R$ are normalized coefficients of \eqref{eq:eigenstates}, the effective modal surface is theoretically given by (see details in the Supplemental Information):
 \begin{equation}
S_{eff}^{triple}=\frac{S_{eff}^{single}}{\max \left\{ {|c_L|^2,|c_M|^2,c_R|^2}\right\}}.
\label{eq:effective_surface}
\end{equation}
Figure \ref{fig6} compares the variations of the effective modal surface as a function of $L/a_0$ for the lossy modes modes 1 and 2 and the quasi-BIC mode 3, with an excellent agreement between the analytical model (\eqref{eq:effective_surface}) and the numerical results obtained from 2D FDTD simulations. The effective modal surface of the lossy mode 1 is always superior to $2S_{eff}^{single}$ except at magic distances where its effective modal surface is exactly equal to that of the lossy mode 2 and to $2S_{eff}^{single}$. Most importantly, at the magic distances, the effective modal surface of the quasi- BIC mode 3 is exactly equivalent to the one of the single-supercell cavity $S_{eff}^{single}$. This effective modal surface tends quickly to $1.5S_{eff}^{single}$ when the distance $L$ is set out of magic configurations. These results demonstrate convincingly the unconventional localization of light in the middle supercell.  
As discussed previously, the effect of magic configurations on the quasi-BIC mode 3 is two fold: i) boosting of quality factor to form a quasi-BIC and ii) reduction of the spatial confinement to  one moiré supercell. It thus offers an ideal design to enhance Purcell effect where light is tightly confined within few micro-size cavity of extremely low losses.  We define a figure of merit to estimate the Purcell enhancement factor of the quasi-BIC mode 3 in the triple-supercell cavity with respect to the single-supercell cavity, given by $\rho=\frac{Q^{triple}/S^{triple}_{eff}}{Q^{single}/S^{single}_{eff}}$. The variations of $\rho$ are represented figure \ref{fig7}, showing again a perfect agreement between the analytical model and the numerical simulations. Impressively, the results show that the Purcell enhancement goes from 1 to 1000 when the distance $L$ is tuned to a magic distance. Finally, we note that although our design is a 2D structure without any confinement along z direction, 3D structures can be harnessed by using rods of finite length for the gratings. It is then possible to engineer a 3D confinement of light with high nominal Purcell factor at magic configurations to enhance light-matter interactions.

\section{Conclusion}

Theoretical investigations have been conducted on microcavities, composed of one or several supercells of a 1D moiré structure. The optical properties of these resonators are accurately described by an analytical model, based on a non-Hermitian Hamiltonian, which has been fully validated through ab initio simulations (FEM, FDTD). It has been demonstrated that, for a microcavity composed of a moiré triple-supercell, the “magic configuration” leads to the emergence of a quasi-bound state with Q-factors exceeding $10^6$, confining light  completely  within the central supercell. As a result, it is shown that a minimum of three moiré supercells is required to achieve unconventional light localization in a finite moiré cavity. This subsequently results in a thousand-fold increase in the Purcell enhancement factor compared to a microcavity composed of a single supercell.

While our primary focus is on the 1D moiré structures, similar effects are anticipated in 2D moiré structures\,\cite{Dong2021,Tang2021}. Generally, to achieve unconventional localization at magic configurations in a single-moiré supercell of finite structures,  it is necessary to surround the supercell with additional supercells. These additional supercells replicate the same lateral interface required for light tunneling, similar to those observed in infinite moiré structures. For instance, to observe unconventional localization at magic angles within finite cavities of a triangular lattice, a minimum of six moiré supercells is expected.

In conclusion, our research uncovers a novel mechanism for enhancing light-matter interaction in subwavelength photonic structures by utilizing moiré cavities that operate at these magic configurations. This could present various applications in quantum optics, optical tweezers, sensing, and lasing devices.

{\large\textbf{Acknowledgement}}
The authors thank Xuan Dung Nguyen,
Duy Hoang Minh Nguyen, Serge Mazauric and Pierre Vicktorovitch for fruitful discussions.\\



Supplementary Information is available for this paper.\\ 







\bibliography{main.bib}

\onecolumngrid

\begin{center}
	\textbf{\large --- SUPPLEMENTAL MATERIAL ---}
\end{center}

\setcounter{equation}{0}
\setcounter{figure}{0}
\setcounter{table}{0}
\setcounter{section}{0}
\setcounter{page}{1}

\renewcommand{\theequation}{S\arabic{equation}}
\renewcommand{\thefigure}{S\arabic{figure}}
\renewcommand{\thesection}{S\arabic{section}}
\renewcommand{\bibnumfmt}[1]{[S#1]}
\renewcommand{\vec}[1]{\boldsymbol{#1}}

\section{1D periodic moiré}

\begin{figure}[!ht]
	\begin{center}
	\includegraphics[width=0.8 \textwidth]{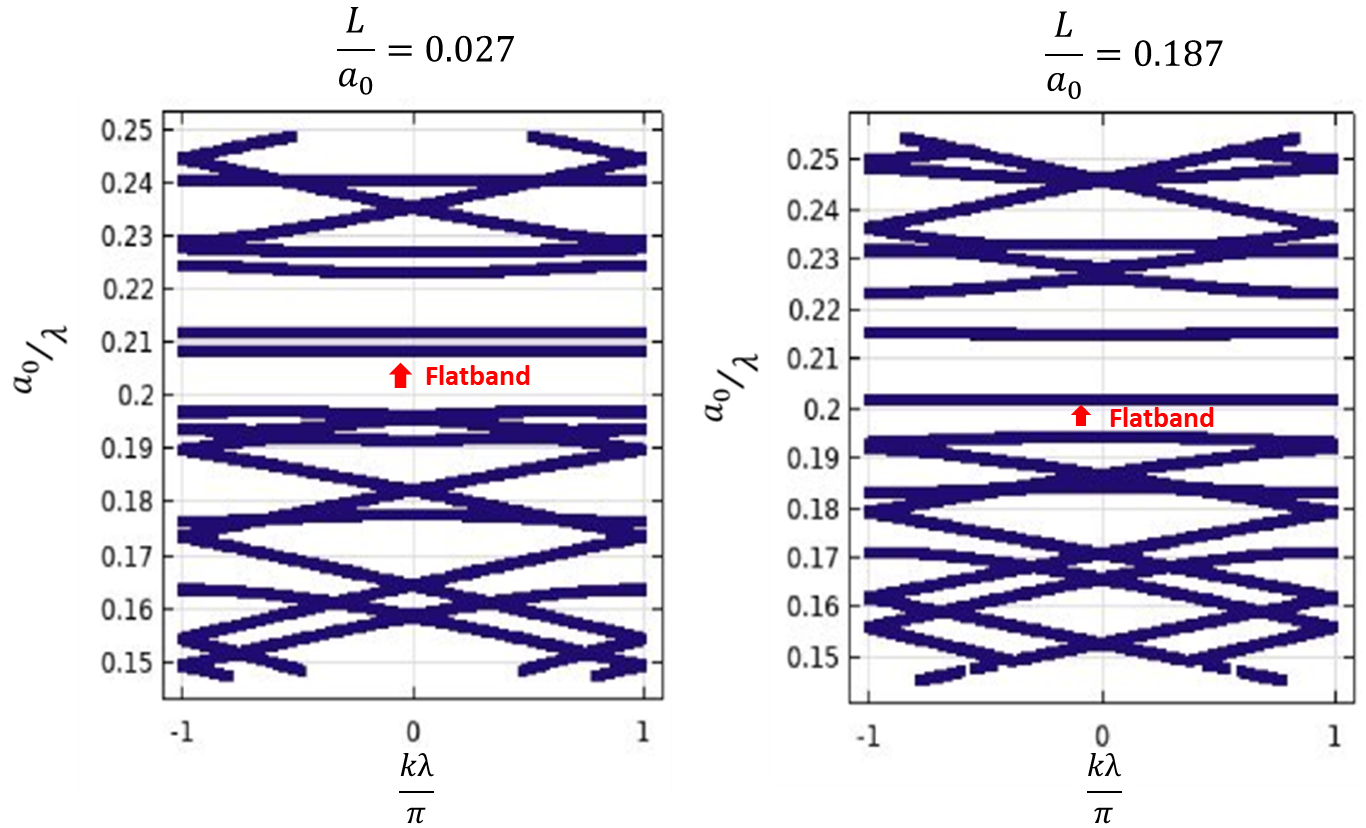}
\caption{Band diagrams corresponding to the magic distances $L/a_0=0.027$ and $L/a_0=0.187$ for 1D moiré.}
	\label{fig:band}
\end{center}
\end{figure}

The band diagrams of 1D moiré are simulated by the finite element method using COMSOL Multiphysics (fig.\ref{fig:band}). We analyze the band structure of the lower flatband frequency.

\begin{figure}[!ht]
\begin{center}
	\includegraphics[width=0.8 \textwidth]{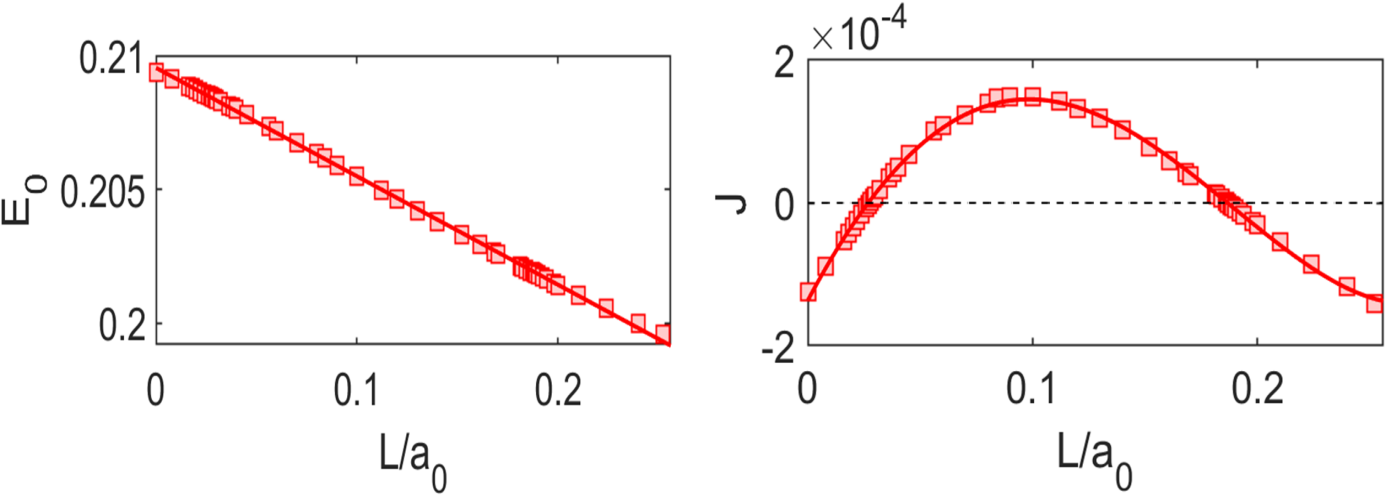}
\caption{Variation of the average energy $E_0$ of 1D moiré, and the tunneling rate $J$ in function of $L/a_0$.}
	\label{fig:ENERGY}
\end{center}
\end{figure}
We compute the energy at $\Lambda (k=0)$ and  $X (k=1)$ sites of the brillouin zone for the lower flatband frequency in order to estimate, from eq.\ref{eq:1}, the average energy $E_0$ and the tunneling rate $J$ given by:
\begin{equation}
    E_0=\frac{E_\Lambda+E_X}{2}
\end{equation}
and
\begin{equation}
 J=\frac{E_\Lambda-E_X}{4}
\end{equation}
\begin{figure}[!ht]
	\begin{center}
	\includegraphics[width=0.8 \textwidth]{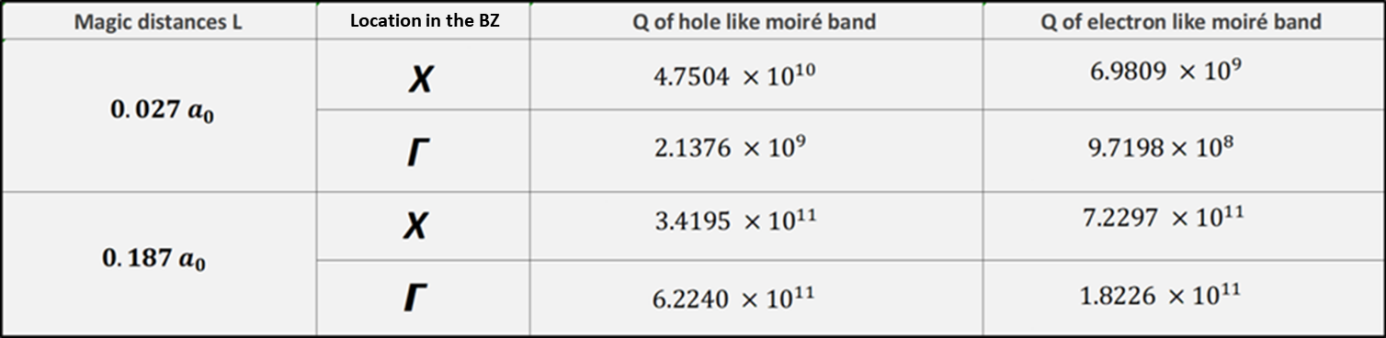}
\caption{ Quality factors of 1D moiré for the magic distances $L/a_0=0.027$ and $L/a_0=0.187$ at BZ sites.}
	\label{TABLE}
\end{center}
\end{figure}
    
  For the two magic distances $L/a_0=0.027$ and $L/a_0=0.187$, the quality factors of the 1D moiré of the hole-like moiré band and the electron-like moiré band at high symmetry points of the BZ sites are shown in the table (\ref{TABLE}). 
\section{Double-supercell microcavity}
We study a microcavity which is made of two supercells known as “diatomic molecules”. Each atom has a defined energy  and is coupled to the other by a tunneling rate  (see fig.\ref{fig:ENERGY}). At the extremities, each atom has a loss  as defined in the case of a single atom. Using as basis, the eigenmodes of the system are described by the following Hamiltonian:

\begin{equation} H=
\begin{pmatrix}
   E_0+\Delta  & J\\
    J & E_0+\Delta
\end{pmatrix}
+i \begin{pmatrix}
   \gamma_0  & 0 \\
   0 & \gamma_0
\end{pmatrix}
\label{eq:H1}
\end{equation}

From the Hamiltonian, we compute the eigenvalues of the system by:

\begin{equation}
    E_{1,2}=E_0+\Delta \pm J+i\gamma_0
\end{equation}

and the eigenvectors by:

\begin{equation}
    \ket{\Psi_{1-2}}= \frac{\ket{\Psi_L}+\ket{\Psi_R}}{2}
\end{equation}

\begin{figure}[!ht]
\begin{center}
	\includegraphics[width=0.7 \textwidth]{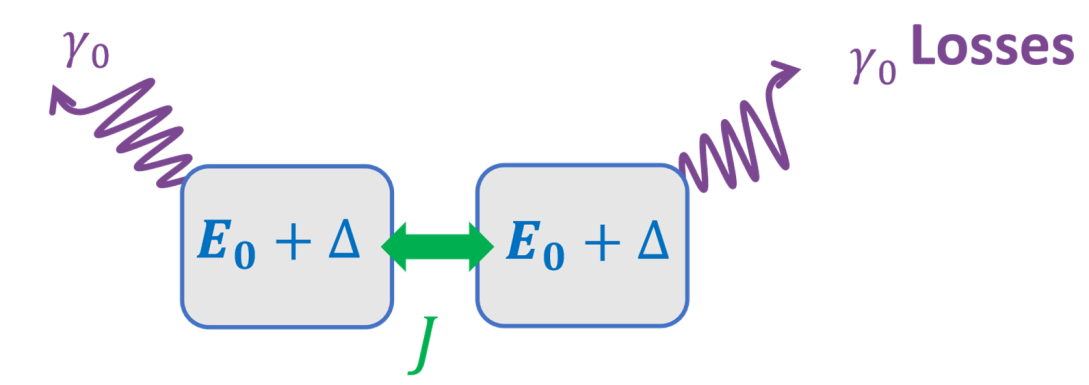}
\caption{Variation of the average energy $E_0$ of 1D moiré, and the tunneling rate $J$ in function of $L/a_0$.}
	\label{fig:ATOMS}
\end{center}
\end{figure}

Figures (S.\ref{fig:ATOMS}.a) and (S.\ref{fig:ATOMS}.b) compare the energy and losses of the two states when scanning the distance obtained by an analytical model to the numerical simulations. The results of the analytical model and the numerical simulations match perfectly.\newpage 
\begin{figure}[ht!]
\begin{center}
	\includegraphics[width=0.8\textwidth]{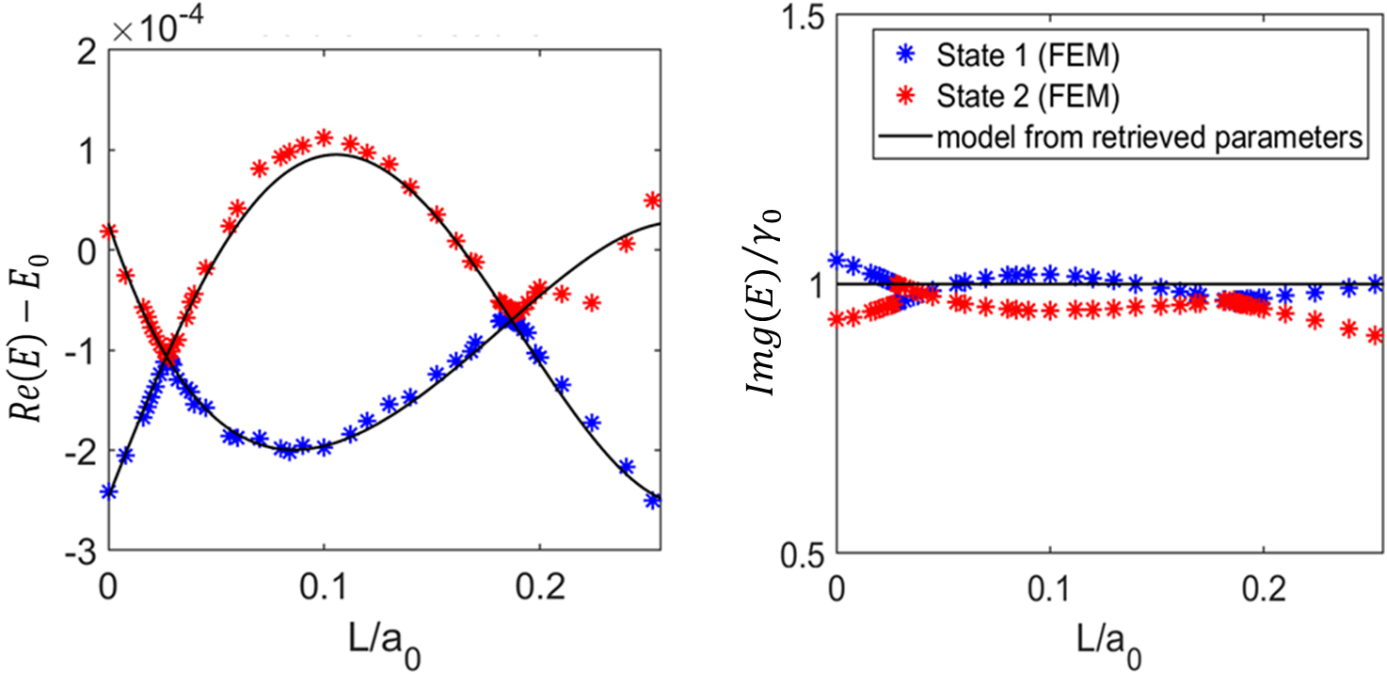}
\caption{(a) The energy of the two bound states, and (b) the losses through the structure of the diatomic molecule as a function of $L/a_0$. Blue and red points are results from finite element method simulations. The black lines are analytical calculations.}
	\label{fig:EJ}
\end{center}
\end{figure}
The model predicts the splitting of the hole-like bound state into bonding and antibonding states separated by . As expected, this splitting vanishes at magic distances (see fig.\ref{fig:EJ}).

\section{Derivation of losses from the eigenvectors}

In the main text, we showed that the losses are given by the imaginary part of the complex eigenvalues. We can also derive the losses from the eigenvectors as an alternative method.
We derive the losses from the normalized eigenvector defined as:
\begin{equation}
    \ket{\Psi}= c_L\ket{L}+c_M\ket{M}+c_R\ket{R}  
\end{equation}

Here, the left and the right atoms which are at the extremities are the origin of the losses. Therefore, from the model, the losses are given by:

\begin{equation}
    \frac{\gamma}{\gamma_0}=\left|c_L\right|^2+\left|c_M\right|^2=2\left|c_L\right|^2
\end{equation}

\begin{figure}[ht!]
\begin{center}
	\includegraphics[width=0.8\textwidth]{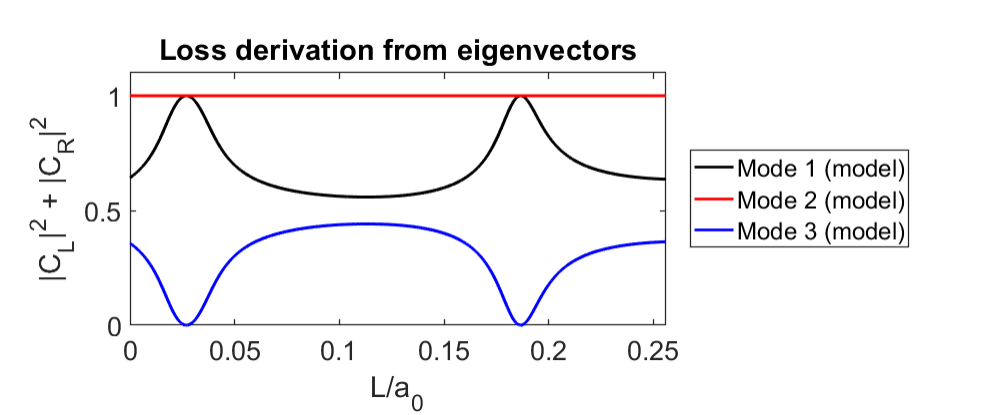}
\caption{Variation of the loss that is derived from the eigenvectors in function of  for the three modes of the triple-supercell cavity.}
	\label{fig:LOSS}
\end{center}
\end{figure}
This method is in good agreement with the method using the imaginary part of the eigenvalues showed in the main text (see fig.\ref{fig:LOSS}).

\section{Derivation of the effective surface from the eigenvectors}

Since $\ket{\Psi}=c_L\ket{L}+c_M\ket{M}+c_R\ket{R}$, the electric field distribution of the triple atom cavity is given by:
\begin{equation}
    F_{triple}\left(x,y\right)=c_LF_{single}\left(x-\mathrm{\Lambda},y\right)+c_MF_{single}\left(x,y\right)+c_RF_{single}\left(x+\mathrm{\Lambda},y\right) 
\end{equation}
   
where $F_{single}\left(x,y\right)$ is the field distribution for the single supercell cavity that is localized around x=0.

Then, we can calculate the following function:
\begin{align}
\begin{split}
F_{triple}\left(x,y\right).n_{triple}\left(x,y\right) 
&=\left[c_LF_{single}\left(x-\mathrm{\Lambda},y\right)+c_MF_{single}\left(x,y\right)+{\ c}_RF_{single}\left(x+\mathrm{\Lambda},y\right)\right].n_{triple}\left(x,y\right) \\
&=c_LF_{single}\left(x-\mathrm{\Lambda},y\right).n_{single}\left(x-\mathrm{\Lambda},y\right)+c_MF_{single}\left(x,y\right).n_{single}\left(x,y\right)\\
&+{\ c}_RF_{single}\left(x+\mathrm{\Lambda},y\right).n_{single}\left(x+\mathrm{\Lambda},y\right)
 \end{split}
 \end{align}
where $n_{single}(x,y)$ and $n_{triple}(x,y)$ are the refractive index distribution for the single and triple cavity respectively.
The surface integral of the function  $\left|F_{triple}\left(x,y\right).n_{triple}\left(x,y\right)\right|^2 $over x and y is defined by:

\begin{align}
    \begin{split}
\iint{\left|F_{triple}\left(x,y\right).n_{triple}\left(x,y\right)\right|^2dxdy}
&=(\left|c_L\right|^2+\left|c_M\right|^2+\left|c_R\right|^2)\iint{\left|F_{single}\left(x,y\right).n_{single}\left(x,y\right)\right|^2dxdy}\\
&=\iint{\left|F_{single}\left(x,y\right).n_{single}\left(x,y\right)\right|^2dxdy})
    \end{split}
\end{align}

then,
\begin{align}
    \begin{split}
\max{\left(\left|F_{triple}\left(x,y\right).n_{triple}\left(x,y\right)\right|^2\right)}&=\max{\left|c_L\right|^2,\left|c_M\right|^2,\left|c_R\right|^2}.\max{\left(\left|F_{single}\left(x,y\right).n_{single}\left(x,y\right)\right|^2\right)}\\
 &=\max{\left\{\left|c_L\right|^2,\left|c_M\right|^2\right\}}.max{\left(\left|F_{single}\left(x,y\right).n_{single}\left(x,y\right)\right|^2\right)} 
    \end{split}
\end{align}

We define the effective surface for the triple atom cavity by;
\begin{align}
\begin{split}
S_{eff}^{triple}&=\frac{\iint{\left|F_{triple}\left(x,y\right).n_{triple}\left(x,y\right)\right|^2dxdy}}{\max{\left(\left|F_{triple}\left(x,y\right).n_{triple}\left(x,y\right)\right|^2\right)}}\\
&=\frac{1}{\max{\left\{\left|c_L\right|^2,\left|c_M\right|^2\right\}}}.\frac{\iint{\left|F_{single}\left(x,y\right).n_{single}\left(x,y\right)\right|^2dxdy}}{max\left(\left|F_{single}\left(x,y\right).n_{single}\left(x,y\right)\right|^2\right)}\\
&=\frac{S_{eff}^{single}}{\max{\left\{\left|c_L\right|^2,\left|c_M\right|^2\right\}}}
\end{split}  
\end{align}
therefore, 
\begin{equation}
    \frac{S_{eff}^{triple}}{S_{eff}^{single}}=\frac{1}{\max{\left\{\left|c_L\right|^2,\left|c_M\right|^2\right\}}}
\end{equation}
\newpage
\begin{figure}
    \centering
    \includegraphics[width=0.8 \textwidth]{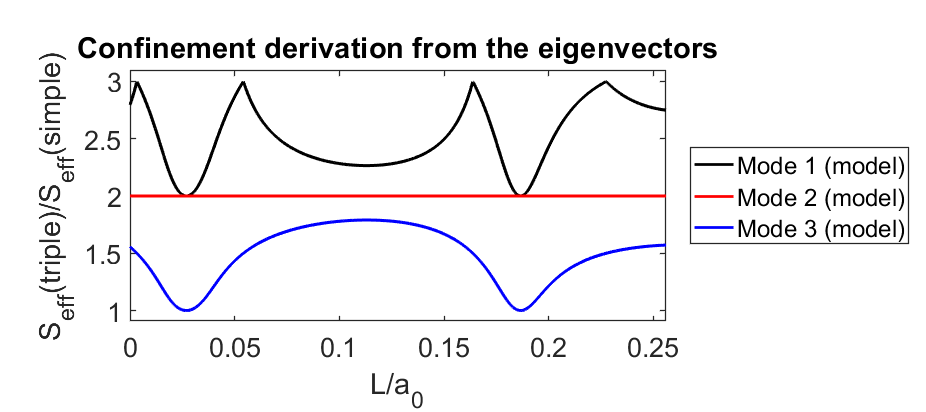}
    \caption{Variation of the effective surface of the triple atom cavity over the effective surface of the single atom cavity in function of $L/a_0$}
    \label{SURFACE}
\end{figure}

Figure \ref{SURFACE} depicts the confinement derivation from the eigenvectors, which agrees well with FDTD simulations showed in the main text.

\end{document}